_{?}
\RequirePackage{fix-cm}
\documentclass{svjour3}                     
\smartqed  
\usepackage{graphicx}
\journalname{General Relativity and Gravitation}
\begin{document}

\title{Testing two alternatives theories to dark matter with the Milky Way dynamics}


\titlerunning{Testing alternatives to dark matter}        

\author{P.L.C. de Oliveira  \and J.A. de Freitas Pacheco \and G. Reinisch}
        

\institute{P.L.C de Oliveira \at
              Departamento de F\'isica, Universidade Federal do Espirito Santo\\
              29075-910  Vit\'oria ES Brazil\\
              \email{paulo\textunderscore lco@yahoo.com.br}           
           \and
           J.A. de Freitas Pacheco and G. Reinisch \at
            Observatoire de la Cote d'Azur\\
							06304 Nice Cedex 4 France\\
							\email{pacheco@oca.eu, gilbert.reinisch@oca.eu}}

\date{Received: date / Accepted: date}

\maketitle

\begin{abstract}

Two alternative theories to dark matter are investigated by testing their ability to describe
consistently the dynamics of the Milky Way. The first one refers to
a modified gravity theory having a running gravitational constant and the second assumes
that dark matter halos are constituted by a Bose-Einstein condensation. The parameters
of each model as well as those characterizing the stellar subsystems of the Galaxy were
estimated by fitting the rotation curve of the Milky Way. Then, using these parameters,
the vertical acceleration profile at the solar position was computed and compared
with observations. The modified gravity theory overestimates the vertical acceleration derived from
stellar kinematics while predictions of the Bose-Einstein condensation halo model 
are barely consistent with observations. However, a dark matter halo based on a collisionless fluid
satisfies our consistency test, being the best model able to describe equally well
the rotation curve and the vertical acceleration of the Galaxy.

\end{abstract}

\section{Introduction}
\label{intro}

Astronomical observations on scales larger than several kiloparsecs indicate that 
the strength of gravitational forces cannot be explained only by the action of
baryonic matter. A component, dubbed dark matter, with a cosmic abundance of about six
times that of baryons and whose nature is still unknown, is necessary to explain different 
data as, for instance: i) the amplitude of peaks observed in the cosmic microwave background
angular power spectrum; ii) the dynamical masses of galaxy clusters, which are larger than 
those under the form of stars or/and hot gas; iii) the gravitational lensing effects produced by massive
galaxies or clusters on more distant objects and iv) the formation of structures in the universe, the 
so-called ``cosmic web". However, in the literature, one of the most discussed effects is 
the ``flat" rotation curve displayed by spiral galaxies that cannot be explained simply by the 
baryonic matter associated to the observed light distribution of those objects.

 Since the expected Standard Model relics do not have abundances sufficiently high to explain the aforementioned
 observations (excepting neutrinos, which have an adequate abundance but are not massive enough and have 
additional difficulties related to their free-streaming length), dark matter particles are expected to be issued
from alternative models. Minimal Supersymmetric extensions of the Standard Model (MSSM) offer a
plethora of candidates like gravitinos, photinos, s-neutrinos among others and presently, the
neutralino, the lightest super-symmetric particle, is one of the most plausible candidates. So far,
no signal of super-symmetry has been seen in experiments performed with the Large Hadron Collider 
\cite{bechtle}. Moreover, results from direct search experiments are controversial.
A positive signal modulated with a period of about one year is claimed to be present in data from 
experiments like DAMA/LIBRA, CoGent and CRESS-II \cite{bernabei,kelso,arina} that
are not confirmed by the more sensitive experiments like XENON100 and LUX \cite{aprile,lux}. Negative results
were also obtained from indirect searches related to the detection of high energy photons, leptons and hadrons
originated from the annihilation of dark matter particles \cite{desai,peirani,jose,lavalle,hatzen,vogl}.
Thus, the only evidence to date for dark matter comes from its gravitational effects and, consequently, the
possibility that General Relativity and the Newtonian theory break down at scales of kiloparsecs
cannot be excluded a priori.

This situation has stimulated the investigation of alternative modified gravitational theories 
in order to explain, in particular, the ``flat" rotation
curve of galaxies. An early proposal in this sense, dating from more than 30 years ago is the MOdified 
Newtonian Dynamics (MOND) \cite{milgrom}. This theory assumes the existence of a critical 
acceleration $a_0 \sim 10^{-8}~cm s^{-2}$. For motions in which the acceleration of bodies is much larger than $a_0$, the gravitational force is essentially Newtonian, while in the opposite regime, the true acceleration is the geometric mean of the Newtonian value and $a_0$. This simple theory is able to explain flat rotation curves and the Tully-Fisher 
relation \cite{gaugh}, as well as some dynamical aspects occurring in the local Universe \cite{kroupa}. However, MOND 
has difficulties to explain the dynamics of galaxy clusters as, for instance, the so-called 
Bullet-Cluster \cite{clowe} and the cosmic matter power spectrum \cite{dodelson}.

Corrections to the Newtonian dynamics is a common feature of different models of effective low-energy quantum theory
of the gravitational field \cite{antoniades,donoghue,bohr}. Based on these ideas, the Renormalization Group
approach was considered by \cite{shapiro,shapiro2,reuter}, in which a ''running'' gravitational constant, depending 
on a given energy scale is expected. The consequences of the possible variation of $G$ in galactic scales, under the assumption
that the energy scale is fixed by the local Newtonian potential was investigated by the authors of reference \cite{ilya}. 
According to them, Renormalization Group effects on General Relativity (RGGR) are able to explain the rotation curve
of disk galaxies with a fit quality better than MOND or the Scalar Tensor Vector Gravity (STVG) \cite{moffat}.
Moreover, this modified gravitation theory fits quite well the observed velocity dispersion profile of elliptical
galaxies \cite{davi}. 

Another possibility often discussed in the literature assumes that dark halos are constituted by
massive scalar fields or bosons that have undergone a Bose-Einstein condensation (BEC) in which
they occupy the same quantum ground level \cite{sin,koh,bohmer,lee,bernal,huang}. Different investigations claim
that the rotation curve of dwarf and low surface brightness (LSB) galaxies can be explained if a BEC halo  
is included in their mass distribution \cite{bohmer,harko,dwornik}.

In the present paper the dynamics of the Galaxy is reviewed using either the RGGR theory or
a dark halo constituted by a BEC. The Newtonian potential, necessary
to fix the energy scale of the ''running'' gravitational constant in the RGGR model, was computed taking 
into account the contribution
of different stellar systems: the central bulge, the thin and the thick disks and also the neutral+ionized gas. The
masses of the stellar components as well as $V_\infty$, a parameter characterizing the effective acceleration
in the RGGR theory, were estimated by fitting the rotation curve of the Milky Way (MW). The same procedure was adopted
for the BEC model. Once
the parameters of these models were fixed by the fit of the rotation curve, the consistency of these theories
was tested by computing and comparing with data, the vertical acceleration at the solar position.  Our calculations indicate that under these conditions the RGGR theory, despite the good description of the
galactic rotation curve, overestimates the vertical acceleration and
this constitutes a potential problem for this model. The fit quality of the rotation curve 
resulting from the BEC model is worse than that of the RGGR theory but the predictions of the
vertical acceleration are barely consistent with data, and alone, they cannot exclude this model.
As we shall see, the difficulties for the BEC model concern the size of the halos 
as well as with the behavior of the rotation curve of the Galaxy beyond 15 kpc. Only models of dark matter
halos constituted by a collisionless fluid and having a mass distribution derived from
cosmological simulations are able to satisfy our consistency test.
This paper is organized as follows:
in Section 2 the main aspects of the RGGR theory is briefly reviewed and the rotation curve of the Galaxy derived from
this approach is discussed; in Section 3 a similar analysis is performed for the BEC dark matter model; in Section 4, for 
a comparison with these models, the same analysis is made for a dark matter halo having a 
Navarro-Frenk-White (NFW) density 
profile; in Section 5 the comparison of all these models with vertical acceleration data is performed and, finally,
in Section 6 the main conclusions of this investigation are given. 

\section{The RGGR theory}

In this theory, quantum corrections produce a running gravitational 
constant $G(\mu)$ that depends now on the energy scale
$\mu$ of the theory (see \cite{ilya,davi} and references therein for details), i.e.,
\begin{equation}
\label{running}
G(\mu) = \frac{G_0}{1+2\nu~ln (\mu/\mu_0)} \approx G_0\left[1-2\nu~ln(\mu/\mu_0)\right]
\end{equation}
where $\nu$ is a small dimensionless parameter. We adopt here the same approach 
as \cite{ilya}, namely, the energy scale $\mu$ is assumed to be related with the
local Newtonian potential by a simple power law or, in other words
\begin{equation}
\label{scale}
\frac{\mu}{\mu_0} = \left(\frac{\phi_N}{\phi_{N,0}}\right)^{\alpha}
\end{equation}
Using the field equations in the weak gravitational regime, it is possible to show that the effective
potential of the theory in this approximation is
\begin{equation}
\label{potential}
\phi_{ef}=\phi_N + \frac{c^2}{2}\frac{\delta G}{G_0}
\end{equation}
where $\phi_N$ is the usual Newtonian gravitational potential. In this case, the effective 
acceleration $K_{ef}$, computed from the gradient of eq.\ref{potential} combined with eqs.\ref{running} 
and \ref{scale}, is given by
\begin{equation}
\label{effacceleration}
K_{ef,i}= -\frac{\partial\phi_N}{\partial x^i} + \frac{\alpha\nu c^2}{\phi_N}\frac{\partial\phi_N}{\partial x^i}
\end{equation}
Since the expected circular velocity, assumed to be equal to the rotation velocity, is given by
\begin{equation}
\label{Vcircular1}
V^2_c =  r\frac{\partial\phi_{ef}}{\partial r}=rK_{ef,r}
\end{equation}
one obtains from the equation above and eq.\ref{effacceleration}
\begin{equation}
\label{Vcircular2}
V^2_c = V^2_N\left(1-\frac{\alpha\nu c^2}{\phi_N}\right)=V^2_N\left(1-\frac{V^2_\infty}{\phi_N}\right)
\end{equation}
where $V_N$ is the Newtonian circular velocity and we have defined $V^2_\infty = \alpha\nu c^2$. Note that the effective circular velocity depends on the
Newtonian value corrected by a factor that depends on the local Newtonian potential and on the unique RGGR parameter
$V_\infty$, which incorporates the product of the parameters, $\alpha$ and $\nu$.

\subsection{Application to the Milky Way}

The Galaxy is constituted by different sub-systems, which have their own mass distribution and kinematics. In 
order to evaluate the local Newtonian potential required to compute the effective circular velocity 
from eq.\ref{Vcircular2}, the following sub-systems were considered: the neutral+ionized gas, the bulge, the thin 
and the thick disks.

\subsubsection{The gas}

The projected density profile of the gas $\Sigma_g(r)$ was taken from reference \cite{olling}. Since only the projected density profile is given, we assumed that the gas is distributed exponentially along the z-axis with a total scale of height $2H_g = 200 pc$.
In this case, the gas density at the (cylindrical) coordinates $r,z$ is given by
\begin{equation}
\label{gasdensity}
\rho_g(r,z) = \frac{\Sigma_g(r)}{2H_g}e^{-z/H_g}
\end{equation}
and the solution of the Poisson equation for an axisymmetric system, after integrating over
the angular variable is   
\begin{equation}
\label{gaspotential}
\phi_g(R,Z)=-\frac{4G}{H_g}\int_0^{\infty}r\Sigma_g(r)dr\int_0^{\infty}dz \frac{K(u(r,z))}
{\sqrt{R^2+r^2+2rR+(Z-z)^2}}e^{-z/H_g}
\end{equation}
In the equation above $K(x)$ is the elliptic integral of first kind and the function $u(r,z)$ is given by
\begin{equation}
u(r,z) = \frac{4rR}{\left[R^2+r^2+2rR+(Z-z)^2\right]}
\end{equation}
Note that in the numerical computations, the values of the projected gas density given by \cite{olling} were multiplied
by a factor $1.4$ to take into account the presence of helium and trace elements. In this case, integrating the projected
gas density given by \cite{olling} and correcting by such a factor, the total gas mass in the
Galaxy is $9.6 \times 10^9~M_\odot$.

\subsubsection{The bulge}

For the bulge, we have adopted the potential given by \cite{cowsik}, i.e.,
\begin{equation}
\label{bulgepotential}
\phi_b(r,z) = -\frac{GM_b}{(r^2+z^2+b^2)^{1/2}}
\end{equation}
with $b = 0.258~kpc$. Fitting the very inner rotation curve of the MW and neglecting the possible dark matter
contribution, the authors of reference \cite{cowsik} estimated the mass of the bulge to be
$M_b = 1.02\times 10^{10} M_\odot$. In our computations, we allowed the bulge mass to vary around this value
when searching for the best fit of the rotation curve of the MW.

\subsubsection{The disk components} 

The thin and the thick disks were supposed to have a double exponential mass distribution (radially and
vertically) given by
\begin{equation}
\label{diskdensity}
\rho_d(r,z)=\frac{M_d}{4\pi R^2_dH_d}e^{-r/R_d}e^{-z/H_d}
\end{equation}
In the equation above, $M_d$, $R_d$ and $H_d$ are respectively the disk mass, the radial length scale
and the scale of height of the considered component (thin or thick). The potential for a double 
exponential mass distribution was taken from reference \cite{gilmore} and, after some algebra is 
given by the expression
\begin{equation}
\label{diskpotential}
\phi_d(R,Z)=-\frac{GM_d}{R}\int_0^{\infty}\frac{dxJ_0(x)}{\left(1+\frac{R_d^2x^2}{R^2}\right)^{3/2}}
\frac{\left[e^{-Zx/R}-\left(\frac{xH_d}{R}\right)e^{-Z/H_d}\right]}{\left(1-\frac{H_d^2x^2}{R^2}\right)}
\end{equation}
where $J_0(x)$ is the Bessel function of order zero.

In our computations, the different length scales were kept fixed and taken from reference \cite{polido}, who
have modeled the Galaxy using star counts in the $J$, $H$, and $K_S$ bands derived from the
2MASS survey. These scales are respectively $R_d$ = 2.12 kpc, $H_d$ = 0.205 kpc for the thin disk and
$R_d$ = 3.05 kpc and $H_d$ = 0.64 kpc for the thick disk. 

In our fitting procedure of the rotation curve, the masses of both components were allowed to vary but not
independently, since a constraint was imposed. In fact, the projected stellar mass density $\Sigma_*$ at the 
solar neighborhood ($r_\odot$ = 8.3 kpc) is about 36 $M_\odot pc^{-2}$ \cite{garbari}. This value is essentially due 
to the contribution of both disk
components (Note that including the gas contribution, the total baryonic projected density at the solar position
is about 44 $M_\odot pc^{-2}$). Using eq.\ref{diskdensity}, the total projected stellar mass density is
\begin{equation}
\label{vinculo}
\Sigma_*(r_\odot)=7.06\left(\frac{M_{d1}}{10^{10}M_\odot}\right)+11.26\left(\frac{M_{d2}}{10^{10}M_\odot}
\right)~~M_\odot pc^{-2}
\end{equation}
where the subscripts $d1$ and $d2$ refer respectively to the thin and thick disks. Thus, all the models
satisfy the condition above.

\subsection{The rotation curve: RGGR model}

The circular velocity expected in RGGR theory was computed from eq.\ref{Vcircular2}. The total gravitational potential was 
evaluated by summing the contribution of the different components, that is
\begin{equation}
\phi_N = \phi_g + \phi_b + \phi_{d1} + \phi_{d2}
\end{equation}
whereas the Newtonian circular velocity was derived from
\begin{equation}
\label{vrotacao}
V_N^2 = R\frac{\partial\phi_N}{\partial R}
\end{equation}

In the fitting procedure, the masses of the bulge, thin and thick
disks were allowed to vary, as well as the RGGR parameter $V_\infty$ but not the gas distribution.
Our best model is shown in figure 1 where the data points were taken from \cite{vrdata}. These represent
normal points issued from different tracers of the disk kinematics whose data were binned and averaged. 

\begin{figure}
\begin{center}
\rotatebox{-90}{\includegraphics[height=11cm,width=8cm]{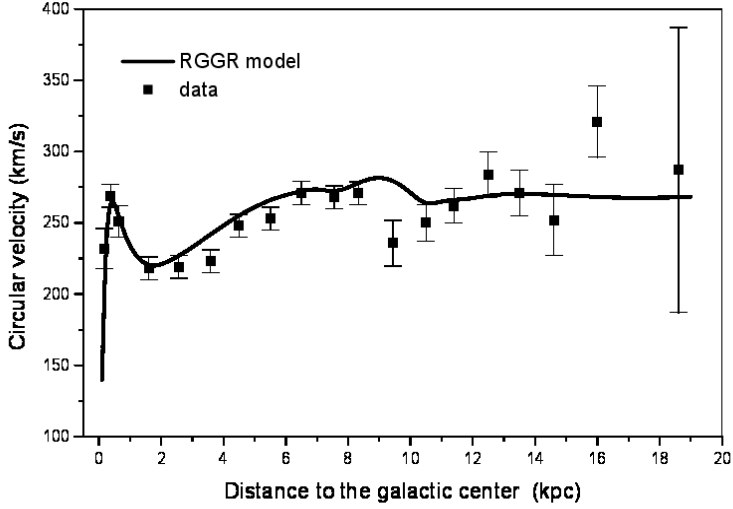}}
\end{center}
\vfill
\caption{Best fit model of the rotation curve of the Galaxy using the RGGR theory. The total
stellar disk mass in this model is $3.58\times 10^{10}~M_\odot$, the bulge mass is $0.78\times 10^{10}
M_\odot$ and the RGGR parameter is $V_\infty = 226~kms^{-1}$.}
\label{fig1}
\end{figure}


This model is characterized by the following parameters: bulge mass = $0.78\times 10^{10}
M_\odot$, thin disk mass = $0.98\times 10^{10} M_\odot$, thick disk mass = $2.60\times
10^{10} M_\odot$ and RGGR parameter $V_\infty$ = 226 $kms^{-1}$. Hence, the resulting
stellar mass of the Galaxy in this model is $M_* = 4.36\times 10^{10} M_\odot$.

The derived RGGR parameter corresponds to $\alpha\nu = 5.67\times 10^{-7}$, which is about
a factor 3.4 higher than that derived from the fit of the rotation curve of NGC 2403 by
the authors of reference \cite{letelier}. They claim that the $\nu$ parameter cannot
vary from galaxy to galaxy but the $\alpha$ parameter can, contrary to MOND or STVG, which
don't have free parameters varying from one object to another. Despite the fact that
the RGGR theory leads to a good fit quality of the rotation curve of the MW, the variation of the 
energy scale among galaxies is a weak point of this theory.

\section{The BEC model}

In this model, dark halos are constituted by an assembly of $N$ identical self-interacting particles of
mass $m$. In the Hartree approach, this self-gravitating structure is dictated by the uncorrelated
single-particle stationary states of the mean-field potential created by the assembly of particles
(Note that in the case of charged particles, the non-linearity of the eigenstates leads to a small
correlation whose amplitude is of the order of the fine structure constant). The
ground state corresponds to a condensed  configuration (BEC) in which all the particles occupy the
lowest-lying orbital of the average potential. The Hartree approximation can be stated as the
one-body self-consistent Schr\"odinger equation, in which the potential energy depends on the
wave-function itself. If $\Phi(r)$ is the single-particle wave-function and $\Psi(r)$ is the
system wave-function in the Hartree sense, then the particle density $n(r)$ is
\begin{equation}
n(r) = \mid\Psi(r)\mid^2 = N\mid\Phi(r)\mid^2
\end{equation}
and the single-particle wave function satisfies the Schr\"odinger equation
\begin{equation}
-\frac{\hbar^2}{2m}\nabla^2\Phi(r) + mU(r)\Phi(r) = E\Phi(r)
\end{equation}
where the mean field potential is defined as
\begin{equation}
U(r) = -GmN\int_{body}d^3r'\frac{\mid\Phi(r')\mid^2}{\mid r-r'\mid} + \frac{4\pi a\hbar^2}{m^2}N\mid\Phi(r)\mid^2
\end{equation}
In the equation above, the first term on the right side represents the gravitational field and the second corresponds
to the ground state of a ''hard-sphere'' potential whose scattering length is $a$. 

When $N \rightarrow \infty$, the solution of these equations can be obtained using the Thomas-Fermi (T-F) approximation, i.e.,
neglecting the so-called quantum correction potential. In this case, the density profile of the configuration 
is simply given by
\begin{equation}
\label{becdensity}
\rho(r) = \rho_0\frac{sin(kr)}{kr}
\end{equation}
The radius $R_{bec}$ of the configuration is obtained with the condition $kR_{bec}=\pi$, corresponding to the 
position $r=R_{bec}$ where the density goes to zero and is given by the relation
\begin{equation}
\label{raiobec}
R_{bec}= \pi\sqrt{\frac{a\hbar^2}{Gm^3}}
\end{equation} 
The exact solution of Schr\"odinger equation, which can be obtained numerically, gives
a density profile that goes only asymptotically to zero when $r \rightarrow \infty$ (the wave function of the
ground state has no nodes). In fact, with respect to the exact solution, $R_{bec}$ includes about 92\% of 
the total mass of the configuration.

The gravitational potential due to the mass distribution given by eq.\ref{becdensity} is
\begin{equation}
\phi_{bec}(r) = -\frac{GM_{bec}}{R_{bec}}\left[1+\left(\frac{R_{bec}}{\pi r}
\right)sin\left(\frac{\pi r}{R_{bec}}\right)\right]
\end{equation}
Note that this equation is valid inside the BEC, i.e., for $r \leq R_{bec}$. When $r > R_{bec}$, the
potential is simply given the point source approximation, this is
\begin{equation}
\phi_{bec} = -\frac{GM_{bec}}{r}
\end{equation}
In these equations, $M_{bec} = 4\rho_0R_{bec}^3/\pi$ is the mass of the BEC in the T-F approximation.
It should be emphasized that is important to distinguish the inner and outer solutions for the BEC
gravitational potential because, as we shall see later, when fitting rotation curves the required BEC radius
is often smaller that typical galactic dimensions.

\begin{figure}
\begin{center}
\rotatebox{-90}{\includegraphics[height=11cm,width=8cm]{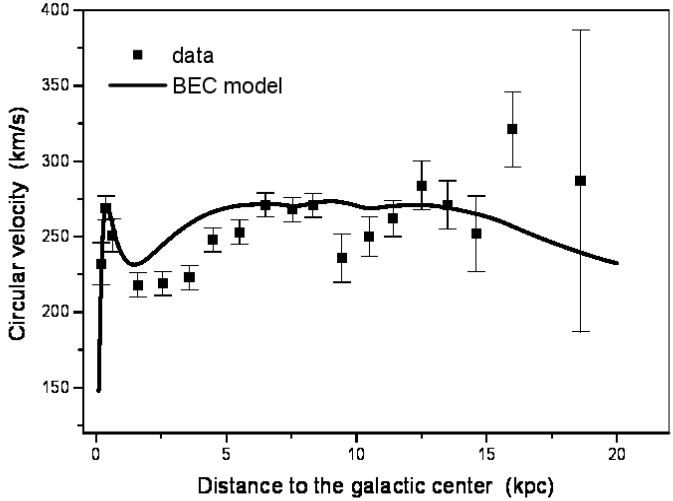}}
\end{center}
\vfill
\caption{Best fit model of the rotation curve of the Galaxy using the BEC model. The total stellar disk
mass in this model is $5.0\times 10^{10}~M_\odot$, the bulge mass is $1.1\times 10^{10}
M_\odot$ and the halo is characterized by a mass of $1.69\times 10^{11}~M_\odot$ and a radius of
15.7 kpc.}
\label{fig2}
\end{figure}


\subsection{The rotation curve: BEC model}

In order to derive the circular velocity, the same procedure as before was used. Firstly,
the total gravitational potential was computed, including all the MW components, i.e.,
\begin{equation}
\label{totalbec}
\phi_N = \phi_{bec} + \phi_g + \phi_b + \phi_{d1} + \phi_{d2}
\end{equation}
where the same notation for the potential of the different galactic systems was used. Then,
in a second step, the circular velocity in the galactic plane was computed from eq.\ref{vrotacao}.
Again, the masses of the stellar components were allowed to vary in the fitting procedure (observed
the constraint mentioned before) as well as the BEC parameters. Note that a unique parameter appears in the 
RGGR theory but two parameters are required for the BEC model: the central density $\rho_0$ (or the total mass) and 
the radius $R_{bec}$. 

The best fit of the rotation curve was obtained for a model in which the masses of the disk components are
respectively $4.79\times 10^{10}~M_\odot$ for the thin disk and $0.21\times 10^{10}~M_\odot$ for the thick
disk. The bulge mass in this model is $1.1\times 10^{10}~M_\odot$. The BEC-halo has the following parameters:
a central density $\rho_0 = 3.43\times 10^7~M_{\odot}kpc^{-3}$, a radius $R_{bec} = 15.7~kpc$ and
a mass of $1.69\times 10^{11}~M_\odot$. 

The first point to be discussed is that in the optimization process of the rotation curve an additional
constraint was necessary, that is the total disk mass was fixed with the value of $5.0\times 10^{10}~M_\odot$.
This was necessary in order to have positive and non-zero values for the masses of both disk components.
As a consequence, our best BEC model has a thin disk much more massive than the thick component. This is
certainly a unrealistic result that could be avoided if the condition expressed by eq.\ref{vinculo} is relaxed.
The second point refers to the radius of the BEC-halo, which is
only 15.7 kpc, whereas galaxy-sized halos constituted by WIMPs are expected to extend up to 200-300 kpc.
In fact, the analysis of the rotation curve of eight dwarf galaxies led to values for the BEC radius down to
1.0 kpc up to 12.6 kpc \cite{harko}. Although that investigation was focused on dwarf galaxies, it is interesting
to recall that the BEC radius does not depend on the mass of the configuration (see eq.\ref{raiobec}) but only
on the scattering length and on the particle mass. Hence, we would expect that all BEC-halos should have similar
dimensions but this is not the case. The third point concerns the derived mass of the BEC halo for the Galaxy.
The mass of Galaxy (mostly in its dark halo) is still uncertain since values for its total mass included
inside a radius of 100 - 200 kpc may vary by factors of 3 to 4. In the one hand, using motions either of the Magellanic 
Clouds and of the Magellanic Stream \cite{lin} or the Sagittarius Stream \cite{gibbons} values around  
$5.5\times 10^{11}~M_\odot$ are obtained. On the other hand, the motion of the satellites of the Milky Way 
indicates higher values that are of the order of $2.0\times 10^{12}~M_\odot$ \cite{sakamoto,kulessa,pacheco}. 
Thus, the derived mass of the BEC halo is about a factor of 3 less than the lower values and almost one order of magnitude less
than the upper values of the estimated mass range. Finally, as it can be seen in figure 2, the modeled circular velocity 
beyond 15 kpc begins to decrease because the gravitational potential of the BEC-halo at these distances varies as $1/r$ as mentioned above. This is in disagreement with observations, which seems to indicate a slightly decline of the rotation
velocity only beyond 60 kpc \cite{vrdata}. All these aspects represent difficulties for this model.

\section{The NFW model}

In this section, the dark halo of the MW is modeled by a NFW profile, derived from fits of halo density profiles 
resulting from numerical simulations, in which dark matter is supposed to behave like a collisionless fluid.
It should be emphasized that the computations presented in this section aim only to serve as a template to be 
compared with the previous models and to check our fitting procedure, since there is an extensive literature on
this subject (see, for instance, \cite{cowsik,nesti} and references therein).

The NFW density profile is given by the relation
\begin{equation}
\label{nfwprofile}
\rho(r) = \rho_c/\left[\frac{r}{R_c}\left(1+\frac{r}{R_c}\right)^2\right]^{-1}
\end{equation}
In the equation above, the halo is defined by two parameters, $\rho_c$ and $R_c$, which are
respectively a characteristic density and radius. The gravitational potential resulting from
the solution of the Poisson equation with the mass density given by eq.\ref{nfwprofile} is
\begin{equation}
\label{nfwpotential}
\phi_{dm}(r)=-4\pi G\rho_cR_c^2\left[\frac{R_c}{r}ln\left(1+\frac{r}{R_c}\right)\right]
\end{equation}

\begin{figure}
\begin{center}
\rotatebox{-90}{\includegraphics[height=11cm,width=8cm]{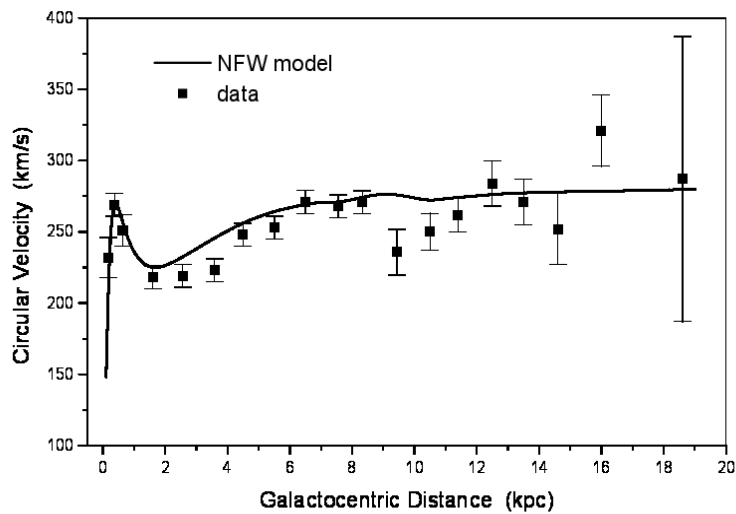}}
\end{center}
\vfill
\caption{Best fit model of the rotation curve of the Galaxy using the NFW model. The total stellar disk
mass in this model is $3.63\times 10^{10}~M_\odot$, the bulge mass is $1.07\times 10^{10}
M_\odot$ and the NFW halo is characterized by the parameters 
$\rho_c = 1.936\times 10^7~M_{\odot}kpc^{-3}$ and $R_c = 17.46~kpc$.}
\label{fig3}
\end{figure}


\subsection{The rotation curve: NFW model}

The same procedure as before was adopted to compute the circular velocity, excepting that in eq.\ref{totalbec}
the BEC potential was replaced by the NFW potential given by eq.\ref{nfwpotential}. The best rotation curve
was searched by varying, as previously, the mass of the stellar components and the two parameters defining
the NFW halo. Figure 3 shows the
best fit model characterized by the following parameters: mass of the thin disk = $1.11\times 10^{10}~M_\odot$,
mass of the thick disk = $2.52\times 10^{10}~M_\odot$ and mass of the bulge = $1.07\times 10^{10}~M_\odot$.
The parameters of the dark halo are respectively $\rho_c=1.936\times 10^7~M_{\odot}kpc^{-3}$ and $R_c = 17.46~kpc$. 
These parameters are similar to those derived in reference \cite{nesti}, whose authors performed a similar 
analysis but including
only the bulge and a single disk besides the halo and obtained $\rho_c=1.40\times 10^7~M_{\odot}kpc^{-3}$ and
$R_c = 16.10~kpc$. The present parameters derived from the fitting of the rotation curve indicate that the halo 
mass inside 200 kpc is about $2.1\times 10^{12}~M_\odot$, in agreement with upper bound estimates based on
the dynamics of satellites of the Local Group \cite{sakamoto,kulessa,pacheco}. 

\begin{figure}
\begin{center}
\rotatebox{-90}{\includegraphics[height=11cm,width=8cm]{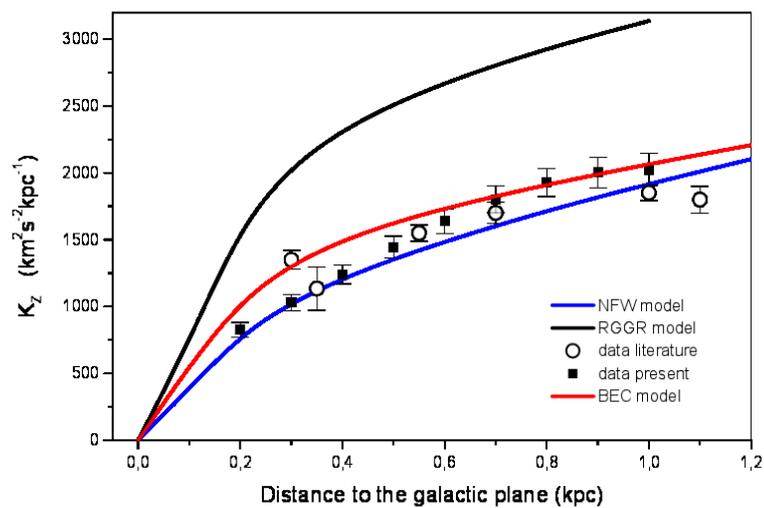}}
\end{center}
\vfill
\caption{Expected vertical acceleration from the considered models compared with data.}
\label{fig4}
\end{figure}


The derived parameters of the MW halo implies that the expected dark matter density in the solar 
neighborhood is $\rho_{dm}(8.3kpc) = 0.75~GeVcm^{-3}$. This value 
is about twice the ''canonical"' value around $0.35~GeVcm^{-3}$ but is consistent with other recent 
independent estimates \cite{cowsik,garbari,weber}.

\section{The vertical acceleration}

In this Section, the consistency of alternative models for dark matter, here represented by a modified
gravity theory (RGGR) and a dark matter particle model (BEC), is tested. Once the parameters
characterizing the gravitational potentials, essentially the masses of the galactic subsystems, and those of the considered model 
are fixed by fitting the rotation curve of the Galaxy, they are expected to give also a 
correct description of the gravitational forces along the vertical axis. 
Thus, using the derived stellar masses for the galactic subsystems and the corresponding parameters, the total
potential can be estimated in any point of the Galaxy. Fixing the radial coordinate at the solar position, this
is $R_\odot = 8.3~kpc$, the vertical acceleration as a function of $Z$ can be computed by the relation
\begin{equation}
\label{zacceleration}
K_Z = -\frac{\partial\phi(R_\odot, Z)}{\partial Z}
\end{equation}

\subsection{The data}

Two sets of data were used in the comparison with expected values of the vertical acceleration shown 
in figure 4. The first is based on the literature, using reference \cite{korchagin} and on the sky 
surveys SEGUE \cite{zhang} and RAVE \cite{rave}. These are indicated by
open circles in figure 4. The second set is based on computations performed using kinematic data on K-dwarfs
taken from reference \cite{garbari}. Their velocity dispersion data as a function of the distance to the galactic
plane was fitted by a second order polynomial, i.e.,
\begin{equation}
\sigma_Z = -16.93 + 16.19Z + 1.11Z^2
\end{equation}
The fit is valid in the range $0.3~kpc \leq Z \leq 1.1~kpc$ and velocities are given in $kms^{-1}$. The density
profile $n_*(Z)$ of K-dwarfs along the vertical axis is given in references \cite{garbari,kuijken} and 
small corrections due to metallicity gradients along the vertical axis were neglected. The vertical acceleration,
neglecting in the Jeans equation small terms coupling the radial and the vertical velocity dispersions, is given by \cite{gilmore}
\begin{equation}
\label{zjeans}
K_Z = \frac{1}{n_*}\frac{\partial(n_*\sigma_Z^2)}{\partial Z}
\end{equation}
Values of the vertical acceleration computed by such a procedure are shown as solid squares in figure 4. 

Simple inspection of figure 4 indicates that the RGGR model overestimates the vertical acceleration. The predicted
acceleration, despite being able to reproduce adequately the effective force field along the galactic plane, fails
to represent the gravitational acceleration along the the vertical axis. Concerning the BEC model,
the predicted rotation curve gives a fit quality worse than that derived for the RGGR (or NFW) model but
the predicted vertical acceleration agrees better with observation than the RGGR theory.

\begin{table}
\caption{Reduced $\chi$-square for fits of the rotation curve and the vertical acceleration.
The first column indicates the model, the second and the third give respectively the reduced
$\chi^2$ and the degree of freedom relative to the fit quality of the rotation curve, while the
last two columns give the same quantities relative to the vertical acceleration.}
\label{tab:1}       
\begin{tabular}{lllll}
\hline\noalign{\smallskip}
Model & $\chi^2_R/\nu$ & $\nu_R$ & $\chi^2_Z/\nu$ & $\nu_Z$  \\
\noalign{\smallskip}\hline\noalign{\smallskip}
RGGR & 1.83 & 15 & - & - \\
BEC & 4.61 & 13 & 10.09 & 14 \\
NFW & 2.17 & 14 & 5.27 & 14 \\
\noalign{\smallskip}\hline
\end{tabular}
\end{table}
 
Table 1 summarizes our results giving the score of these different models measured by the reduced $\chi$-square,
related either with the description of the rotation curve or the vertical acceleration profile.
It is worth mentioning that the relative high values of the reduced ``$\chi^2$'' associated
to the analysis of the vertical acceleration are due only to one or two points whose observational
errors were probably underestimated.
The RGGR theory gives a description of the rotation curve of the MW slightly better than the NFW
profile but overestimates largely the vertical acceleration. The BEC halo model gives the worse
description of the rotation curve but represents the vertical acceleration better than the RGGR
theory. Only the dark matter halo modeled by a collisionless fluid and a NFW density profile gives
an adequate description of both the rotation curve and the vertical acceleration with the same set
of parameters.

\section{Conclusions}

In this work, the consistency of two alternatives to dark matter was investigated. The first one concerns
a modification of gravity, characterized by a running gravitational constant (RGGR theory). The second 
one considers a model for galactic halos, which would be constituted by massive
bosons forming a Bose-Einstein condensation.

In both cases, the rotation curve of the Galaxy served as a departure point to fix the parameters of 
these two theories as well as those defining the different stellar sub-systems of the Milky Way 
like the bulge, the thick and the thin disks. 

Once the different parameters were fixed by the fit of the rotation curve, the consistency of these two
models was tested by computing the expected vertical acceleration at the solar position. As it was shown,
the predicted vertical acceleration profile by the RGGR theory overestimates the data derived from 
different sky surveys and this represents a potential problem for this model.
Moreover, in the RGGR approach the variation of the ''$\alpha$'' parameter from one object to another seems to be
necessary in order to explain the rotation curve of different galaxies, being an additional problem for this
model. Our best BEC halo model gives a representation of the rotation curve of the MW worse than that
obtained by the RGGR theory but the predicted vertical acceleration profile is barely consistent with data.
However, the BEC model has also other difficulties. The radius of the BEC halo varies from galaxy 
to galaxy, contrary to the expectation of the theory since 
such a quantity does not depend on the mass of the considered object but only on the particle mass and 
on the scattering length. Moreover, in the case of the Galaxy, even taking into account 
the uncertainties in the present determination of its mass, the resulting value for the BEC halo
is even smaller than the lower bounds. 

The usual dark matter model, in which halos are constituted by a collisionless fluid and whose density distribution is
derived from numerical simulations, explains consistently both the rotation curve
and the vertical acceleration in the solar neighborhood. The halo mass inside 200 kpc derived from the fit 
of the rotation curve agrees with previous analyses and with other 
independent estimates based on the dynamics of the Local Group. The expected local dark matter density is about a
factor two higher than the usual value of $0.3~GeVcm^{-3}$, but in agreement with other recent independent 
determinations. The present study indicates for the total baryonic mass of the Galaxy (bulge+disk+gas) the
value of $M_{bar} = 5.66\times 10^{10}~M_\odot$ comparable with other recent estimates \cite{newman}.


%
%

\begin{acknowledgements}
P.L.C.O thanks respectively to the Brazilian Agencies Coordena\c c\~ao de Pessoal de N\'ivel Superior (CAPES) and Conselho Nacional de Desenvolvimento Cient\'ifico e Tecnol\'ogico (CNPq) for a PhD fellowship and the financial
support, which has permitted his
one year stay at the Observatoire de la C\^ote d'Azur.
\end{acknowledgements}



\end{document}